\documentclass{article}

\usepackage{arxiv}

\usepackage[utf8]{inputenc} 
\usepackage[T1]{fontenc}    
\usepackage{hyperref}       
\usepackage{url}            
\usepackage{booktabs}       
\usepackage{amsfonts}       
\usepackage{nicefrac}       
\usepackage{microtype}      
\usepackage{lipsum}
\usepackage{graphicx}
\graphicspath{ {./images/} }
\usepackage{cite}
\usepackage{amsmath,amssymb,amsfonts}
\usepackage{graphicx}
\usepackage{multirow}
\usepackage{textcomp}
\usepackage{xcolor}
\usepackage{underscore}
\usepackage{graphicx}
\usepackage{subfigure}
\usepackage{tikz}

\usepackage[linesnumbered,ruled,vlined]{algorithm2e}
\usepackage{algpseudocode}  
\usepackage{amsmath}  
\usepackage{tcolorbox}
\usepackage{mdframed}
\usepackage{float}
\usepackage{booktabs}

\hypersetup{hidelinks}

\usepackage{xurl}

\usepackage{xspace}
\newcommand{\method}{RisConFix\xspace}

\title{RisConFix: LLM-based Automated Repair of Risk-Prone Drone Configurations}

\author{
 Liping Han \\
  School of Computer Science\\
  Nanjing University of Posts and Telecommunications\\
  Nanjing, China \\
  \texttt{liping@njupt.edu.cn} \\
   \And
 Tingting Nie \\
  School of Computer Science\\
  Nanjing University of Posts and Telecommunications\\
  Nanjing, China \\
  \texttt{nietingting0819@gmail.com} \\
  \And
 Le Yu \\
   School of Computer Science\\
  Nanjing University of Posts and Telecommunications\\
  Nanjing, China \\
  \texttt{yulele08@njupt.edu.cn} \\
   \And
 Mingzhe Hu \\
  School of Computer Science\\
  Nanjing University of Posts and Telecommunications\\
  Nanjing, China \\
  \texttt{hmz@njupt.edu.cn} \\
  \And
 Tao Yue \\
   School of Computer Science and Engineering\\
  Beihang University\\
  Beijing, China \\
  \texttt{yuetao@buaa.edu.cn} \\
}

\begin{document}

\maketitle
\begin{abstract}

Flight control software is typically designed with numerous configurable parameters governing multiple functionalities, enabling flexible adaptation to mission diversity and environmental uncertainty. Although developers and manufacturers usually provide recommendations for these parameters to ensure safe and stable operations, certain combinations of parameters with recommended values may still lead to unstable flight behaviors, thereby degrading the drone’s robustness. To this end, we propose a Large Language Model (LLM) based approach for real-time repair of risk-prone configurations (named \method) that degrade drone robustness. \method continuously monitors the drone’s operational state and automatically triggers a repair mechanism once abnormal flight behaviors are detected. The repair mechanism leverages an LLM to analyze relationships between configuration parameters and flight states, and then generates corrective parameter updates to restore flight stability. To ensure the validity of the updated configuration, \method operates as an iterative process; it continuously monitors the drone’s flight state and, if an anomaly persists after applying an update, automatically triggers the next repair cycle. We evaluated \method through a case study of ArduPilot (with 1,421 groups of misconfigurations). Experimental results show that \method achieved a best repair success rate of 97\% and an optimal average number of repairs of 1.17, demonstrating its capability to effectively and efficiently repair risk-prone configurations in real time. 
\end{abstract}

\section{Introduction}
Drones have become widely used in diverse tasks, such as environmental monitoring and emergency response. Unlike conventional smart devices directly operated by humans, drones are controlled by flight control software that autonomously manages flight operations. To support diverse missions and operate under uncertain environmental conditions, modern flight control software (e.g., PX4\cite{PX42025}, and ArduPilot\cite{ardupilot2025}) provide a large number of configurable parameters governing multiple functionalities, for example, flight control (attitude, position, and power system regulation), mission execution (flight mode), hardware adaptation (sensor calibration, actuator mapping, and communication link configuration) etc. However, recent studies have shown that certain combinations of these parameters may induce undesirable flight behaviors, including unstable oscillations and even crashes. In this paper, we refer to this type of improper configuration as \textit{risk-prone configurations}, which poses a severe threat to drone robustness. For example, inappropriate tuning of attitude controller gains may cause oscillations or delays in response, while suboptimal throttle and pitch limit settings can result in altitude divergence or sudden thrust loss.

To ensure configuration reliability, prior studies have primarily focused on detecting misconfigurations before deployment. For instance, LGDFUZZER \cite{han2022control} and ICSEARCHER \cite{han2024range} employ fuzzing and learning-guided search techniques to detect incorrect configurations. While these approaches have proven effective in detecting potential risks, they remain constrained to offline analysis or pre-deployment testing. In practice, however, it is nearly impossible to exhaustively capture all risk-prone configurations during testing. This limitation highlights the necessity of an online and adaptive mechanism capable of detecting and repairing risk-prone configurations in real time, thereby maintaining robust flight behavior during dynamic operations.

To fill this gap, we propose \method, a Large Language Model (LLM) based automated repair framework for risk-prone drone configurations. \method continuously monitors the drone’s operational state and automatically triggers a repair mechanism once abnormal flight behaviors are detected. The repair mechanism leverages an LLM to generate corrective configuration updates that restore flight stability in real time.
Specifically, \method consists of two main phases: 1) Runtime Anomaly Monitor, which continuously collects flight data (e.g., attitude, velocity, control outputs) and detects abnormal flight behaviors via a wireless communication protocol. 2) LLM-based Repair, which formulates the current configuration and flight state into structured prompts and queries an LLM to generate corrective configuration adjustments aimed at mitigating instability. To ensure the validity of each update, \method operates as an iterative repair loop. It keeps monitoring the drone’s state after a fix is applied, and if anomalies persist, it automatically triggers the next repair cycle.

In summary, this paper makes the following contributions:
\begin{itemize}
    \item We propose \method, an LLM–based method that automatically repairs risk-prone configurations of drone flight control software. 
    \item We evaluate \method with extensive experiments (1421 sets of risk-prone configurations in total), and experimental results show that \method effectively and efficiently repairs risk-prone configurations in real time. 
    \item We share our experiences and lessons learned.
    \end{itemize}

The rest of this article is organized as follows. Section \ref{Sec: method} introduces our proposed method for automatically repairing risk-prone configurations. Section \ref{Sec: evaluation} reports the empirical evaluation of \method. Related work is reviewed in Section \ref{Sec: related work}. Finally, Section \ref{Sec: lessons learned} summarizes key insights and lessons learned from our study, and Section \ref{Sec: conclusion} concludes the paper.

\section{Methodology}\label{Sec: method}

\begin{figure}[ht]
 \centering
 \includegraphics[width=0.7\textwidth]{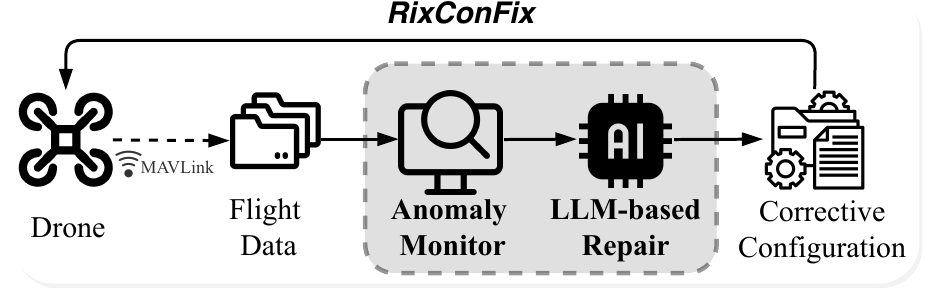}
 \caption{Overview of \method.}
 \label{fig:method-overview}
\end{figure}

\method is designed as an iterative process consisting of two key phases: \textit{Anomaly Monitor}, which continuously monitors flight state; and \textit{LLM-based Repair}, which is triggered upon anomaly detection to repair risk-prone configurations. 
Figure~\ref{fig:method-overview} illustrates the overall workflow of \method, while Algorithm~\ref{alg:repair} details the step-by-step procedures for real-time risk-prone configurations repair. Below, we elaborate on each phase of \method.

\subsection{Anomaly Monitor}
As shown in Algorithm \ref{alg:repair}, to start a flight mission, one must configure the intended functionalities of the flight control software with a set of parameters $P_{initial}$, and upload a mission profile $F_{mission}$ that specifies the intended sequence of waypoints.
During mission execution, \method continuously collects flight data via a wireless communication protocol (i.e., MAVLink) and monitors the incoming data stream to detect whether the drone has entered an anomalous state. 
Below, we summarize the anomaly metrics employed in this study, which were derived from an analysis of the physical effects that different configurations impose on drone behavior \cite{han2024range}:

\begin{itemize}

    \item \textbf{Deviation.} A flight is identified as a \textit{Deviation} when the perpendicular distance (computed using Heron’s formula) between the drone’s position and the planned trajectory exceeds 10 meters for more than 10 consecutive instances. 
    \item \textbf{Thrust Loss.} A \textit{Thrust Loss} is detected when the MAVLink message contains the keyword \textcolor{gray}{\textit{Potential Thrust Loss}}, implying that the flight control software has detected possible propulsion failure or insufficient thrust generation. 
    \item \textbf{Timeout.} A \textit{Timeout} state is detected when the drone remains stationary (or moves only slightly, i.e., its velocity is below 1 meter per second and altitude variation is less than 0.2 meters) for more than 6 consecutive instances.
    \item \textbf{Crash.} A \textit{Crash} is detected when the MAVLink message contains the keywords \textcolor{gray}{\textit{SIM Hit ground}} or \textcolor{gray}{\textit{Crash}}, and the impact velocity exceeds a predefined threshold. Such events indicate that the drone has collided with the ground at an unsafe speed. 
    
\end{itemize}

\subsection{LLM-based Repair} 
As shown in Lines 6-11 of Algorithm \ref{alg:repair}, once an anomaly is detected, the repair mechanism is automatically triggered. 
In Line 7, \method queries the LLM to generate a set of corrective parameter values based on the detected anomaly type $anomaly\_type$ and the current configuration values $P_{current}$. 
It is worth noting that the prompt design also incorporates the official recommended ranges for each parameter\footnote{The recommended range for each parameter is retrieved from the official documentation provided by the flight control software developers.}, which provides the LLM with essential domain knowledge to produce more effective configuration repairs.

\begin{algorithm}  
  \caption{ LLM-based Real-Time Repair }  \label{alg:repair}
  \small
  \KwIn{Initial parameter values $P_{initial}$, Mission profile $F_{mission}$, Predefined repair limit $lm$}  
  \KwOut{Repair record $R$} 
  $ P_{current}\leftarrow P_{initial} $ \\
  $repair\_count \leftarrow 0$  \\
   $monitor \leftarrow StartMission (P_{initial}, F_{mission}) $ \\
\While{$monitor.is\_not\_land( )$ and $repair\_count < lm$}
{
 $anomaly\_type  \leftarrow monitor.monitor\_anomaly()$ \\
\If{$anomaly\_type \neq $ \textnormal{NULL}}
      {
      $P_{fix} \leftarrow repair.get\_fix(anomaly\_type, P_{current})$ \\
      $ repair.upload\_params(P_{fix})$ \\
      $ P_{current} \leftarrow P_{fix}	$ \\
      $repair\_count \leftarrow repair\_count+1	$\\
      $anomaly\_record.append(anomaly\_type)$
       }
}

$result \leftarrow monitor.get\_final\_status()$	\\ 
$StopMission()$\\
$R \leftarrow $ \{$P_{initial}$, $result$, $anomaly\_record$, $repair\_count$\} \\
Return $R$
\end{algorithm}

\begin{figure}[ht]
 \centering
 \includegraphics[width=0.8\textwidth]{Prompt_example.pdf}
 \caption{Shortened example of an LLM prompt. Red text indicates prompt variables, the remaining text forms the static prompt template, and blue text marks the key elements.}
 \label{fig:prompt example}
\end{figure}

\begin{figure}[ht]
 \centering
 \includegraphics[width=0.46\textwidth]{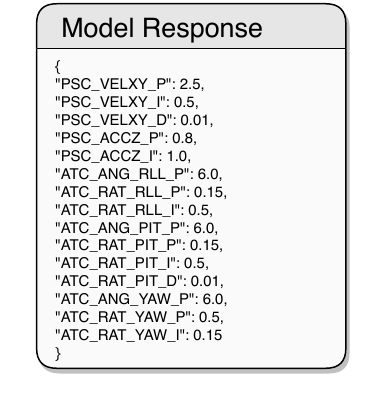}
 \caption{Example model response.}
 \label{fig:response example}
\end{figure}

Figure \ref{fig:prompt example} illustrates a shortened example prompt provided to the LLM. In prompt,
\textcolor{gray}{\textit{\{error_type\}}} represents the detected anomaly category, while \textcolor{gray}{\textit{\{param_info_str\}}} encapsulates the official recommended ranges for each parameter, along with its step size and current configuration values. With this prompt, the LLM infers potential causes of the anomaly and generates corrective configurations (in JSON format), specifying the parameters to be updated and their recommended values (an example response is shown in  Figure \ref{fig:response example}). 
The generated configurations are then uploaded to the flight control software (Line 8), and the \textit{Anomaly Monitor} module continuously validates the effectiveness of the applied updates. If anomalous behaviors persist after reconfiguration, the repair mechanism iteratively triggers subsequent repair cycles. This process continues until the mission is completed, that is $monitor.is\_not\_land( )$ is FALSE, or a predefined repair limit $lm$ is reached (Lines 4-11). 
When the iterative process no longer meets its continuation conditions, the mission is terminated, and the repair outcome is recorded. This record includes whether the mission passed or failed, the anomalies detected, and the total number of repair iterations triggered during the flight (Lines 12–15).

\begin{figure}[htbp]
\centering
\subfigure[Flight trajectory without real-time repair.]{
\includegraphics[width=7cm]{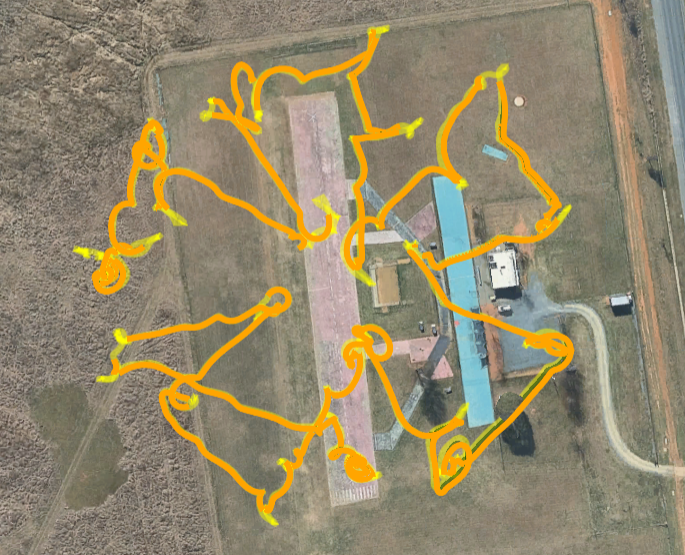}
}
\subfigure[Flight trajectory with real-time repair.]{
\includegraphics[width=7cm]{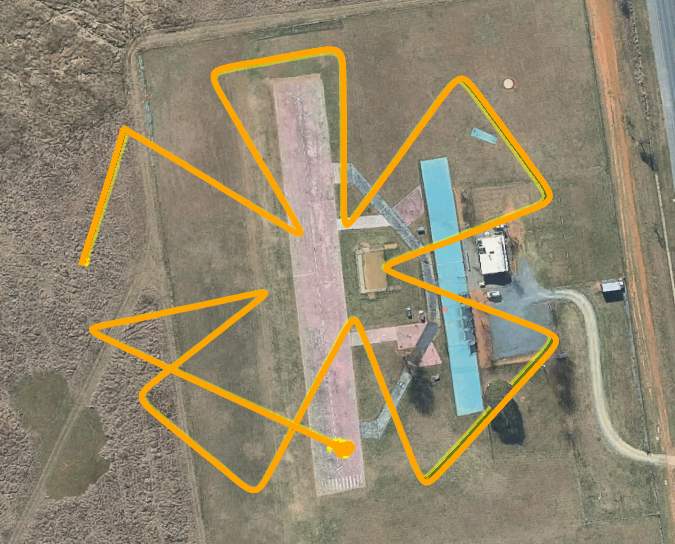}
}
\caption{Comparison of flight trajectories before and after applying the proposed method. 
The upper subfigure illustrates the flight trajectory with the original configuration, while the lower one presents the trajectory after applying the configuration update shown in Figure \ref{fig:response example}. 
}\label{fig:comparison}
\end{figure}

Figure \ref{fig:comparison} shows a comparison before and after the repair operation. 
The upper subfigure shows the flight trajectory with the original configurations, where the drone exhibits noticeable oscillations throughout the mission. In contrast, the lower subfigure depicts the trajectory after applying the response configuration in Figure \ref{fig:response example} upon first detection of an anomaly. Obviously, the drone experienced slight instability at the beginning of the mission (bottom of the subfigure), then the subsequent trajectory became smoother. This demonstrates that the proposed \method can effectively enhance flight robustness in real time.

\section{Empirical Evaluation} \label{Sec: evaluation}

\subsection{Research Questions}
To evaluate \method, we design the following research questions (RQs): 

\textbf{RQ1} - To what extent can \method successfully repair risk-prone configurations that lead to anomalous drone behaviors?

\textbf{RQ2} - How efficient is \method in completing the repair process, as measured by the number of iterations needed to restore stable flight?

\subsection{Evaluation Metrics}
To answer RQ1 and RQ2, we define below two evaluation metrics:

\textbf{Repair Success Rate (RSR)}. $RSR$ quantifies the proportion of cases in which repair operations are triggered and the risk-prone configuration is ultimately repaired successfully.
A case is considered successfully repaired if it meets two criteria: 1) \method actively triggers the repair operation (i.e., generates a targeted fix for the detected anomaly); and
2) The generated configuration(s), when applied, completely resolve the detected anomaly without introducing new anomalies.
The calculation of $RSR$ is as follows: 

\begin{equation}\label{Eq: RSR}
RSR = \left( \frac{NRC}{TTC} \right) \times 100\%
\end{equation}

\noindent where $NRC$ denotes the number of successfully repaired configuration instances, and $TTC$ represents the total number of experimented configuration instances.

\textbf{Average Number of Repairs (ANR)}. $ANR$ quantifies the average number of repair attempts required for each risk-prone configuration to be ultimately repaired successfully. A valid repair attempt is defined as: \method actively triggers a repair operation (i.e., generates a targeted fix for the detected anomaly). For each successfully repaired case, we count the number of repair attempts until a successful repair is achieved.
The calculation of $ANR$ is as follows:

\begin{equation}\label{Eq: ANR}
ANR = \frac{TRA}{NRC}
\end{equation}

\noindent where $TRA$ is the total number of repair attempts for all successfully repaired cases.

\subsection{Experiment Setups and Benchmark Dataset}
To evaluate the effectiveness of \method, we conducted experiments with ArduPilot\cite{ardupilot2025}, 
an open-source drone flight control software that is widely adopted in both commercial and research applications \cite{aliane2024survey}. For our evaluation dataset, we utilized the benchmark dataset 
from \cite{han2025real}, which contains 1,421 distinct misconfigurations. 
Each case in the dataset is a set of potentially risk-prone configurations that may lead to abnormal or unstable flight behaviors. 
To select a suitable LLM for our task, we conducted preliminary experiments comparing multiple state-of-the-art models. 
Based on rigorous pre-experimental evaluation, we selected DeepSeek and Qwen as our models due to their superior performance in generating corrective parameter values. 
For the predefined repair limit, we set $lm=5$. This means that if the number of repair iterations exceeds five, the mission will be marked as a failure, and the anomaly corresponding to the fifth repair attempt will be returned.

\subsection{Experiment Results}

\subsubsection{Effectiveness (RQ1)}

Table \ref{tab:mission_results} summarizes the executed mission results and repair involvement statistics. The mission was executed 1,421 times (i.e., using the 1,421 sets of misconfigurations) on each model.
With DeepSeek, there were 1,419 executions passed and 2 failed. For the passed missions, repair attempts were triggered in 1,380 (1,148+231+1) cases, i.e., 1,148 cases were fixed with one repair attempt, 231 with two attempts, and only 1 mission required three repairs. Additionally, 39 passed missions did not involve any repairs. For the two failed missions, no repair attempts were triggered. We further checked the execution logs and found that the total mission execution times of these two cases surpasses a predefined limit, and therefore failed. 
In summary, the repair success rate $RSR$ is \textbf{97\%} ($1,380/1,421$), indicating that \method can effectively repair risk-prone configurations.

\setlength{\tabcolsep}{12pt}
\begin{table}[htbp]
  \centering
  \caption{Summary of mission results and repair involvement statistics}
  \begin{tabular}{c c c c lc}
    \toprule
   \textbf{LLMs}& \textbf{Results} & \textbf{Counts} & \textbf{Repairs} & \multicolumn{2}{c}{\textbf{Counts}} \\
    \midrule
   \multirow{6}{*}{DeepSeek} & \multirow{4}{*}{Passed}
    & \multirow{4}{*}{1,419} & \multirow{3}{*}{Trigged} & 1 time & 1,148 \\
    \cline{5-6}
    & & & & 2 times & 231 \\
       \cline{5-6}
   & & & & 3 times & 1 \\
      \cline{4-6}
   & & & Not Triggered & \multicolumn{2}{c}{39} \\
     \cline{2-6}
   & \multirow{2}{*}{Failed} & \multirow{2}{*}{2} & Triggered & \multicolumn{2}{c}{0} \\
       \cline{4-6}
  &  & & Not Triggered & \multicolumn{2}{c}{2} \\
    \hline
    
\multirow{8}{*}{Qwen} & \multirow{6}{*}{Passed}
    & \multirow{6}{*}{1,210} & \multirow{5}{*}{Triggered} & 1 time & 309 \\
    \cline{5-6}
    & & & & 2 times & 172 \\
       \cline{5-6}
   & & & & 3 times & 237 \\
      \cline{5-6}
        & & & & 4 times & 214 \\
      \cline{5-6}
        & & & & 5 times & 239 \\
      \cline{5-6}
   & & & Not Triggered & \multicolumn{2}{c}{39} \\
     \cline{2-6}
   & \multirow{2}{*}{Failed} & \multirow{2}{*}{211} & Triggered & \multicolumn{2}{c}{209} \\
       \cline{4-6}
  &  & & Not Triggered & \multicolumn{2}{c}{2} \\
    \bottomrule
  \end{tabular}
  \label{tab:mission_results}
\end{table}

In contrast, for Qwen, it had 1,210 passed missions but a relatively higher number of failed cases (i.e., 211). For the passed missions, repairs were triggered in 1,171 cases (309+172+237+214+239) across multiple repair attempts (up to 5 times), and not triggered in 39 cases. Notably, for failed missions, repairs were triggered in 209 cases, implying that Qwen attempted repairs in most of its failed missions but still could not achieve success. This indicates that the repair effectiveness of Qwen is less favorable compared to DeepSeek.
Nevertheless, we still achieved an \textbf{82\%} ($1,171/1,421$) repair success rate on Qwen.
Overall, \method achieved an average repair success rate of \textbf{90\%} across both models, indicating that \method can effectively repair risk-prone configurations on the fly and substantially enhance flight robustness.

\begin{tcolorbox}[
    colback=gray!10,      
    colframe=gray!80,     
    title=Conclusion for RQ1
]
\method achieved an average repair success rate of \textbf{90\%} across both models, with the best repair success rate reaching \textbf{97\%}, demonstrating the effectiveness of \method.
\end{tcolorbox}

\subsubsection{Efficiency (RQ2)} 

As shown in Table \ref{tab:mission_results}, for the 1,380 successfully repaired cases on DeepSeek, there are a total of 1,613 ($1,148+231\times2+1\times3$) repair attempts, resulting in an average number of repairs of \textbf{1.17} (Eq. \ref{Eq: ANR}). It means that with DeepSeek, \method can fix risk-prone configurations with nearly one repair attempt on average, showing its capability to efficiently identify and repair improper configurations induced anomalies.  
In contrast, for Qwen, across its 1,171 repair-triggered cases within passed missions, the total number of repair attempts accumulates to 3,415 ($309+172\times2+237\times3+214\times4+239\times5$) repair attempts. The corresponding average number of repairs is \textbf{2.91} ($3,415/1,171$), which implies that Qwen requires approximately three repair iterations per triggered case to fix risk-prone configurations. 
Such differences in $ANR$ reflect the models’ robustness in handling risk-prone configurations of flight control software. DeepSeek’s low $ANR$ indicates its reliability in maintaining flight stability with minimal intervention, while Qwen’s relatively higher $ANR$ implies a greater cost in achieving the same level of repair effect. Therefore, choosing a suitable LLM model is crucial for improving repair efficiency. Nevertheless, our method still achieved an average of \textbf{2.04} repairs across both models, demonstrating \method's practicality.

\begin{tcolorbox}[
    colback=gray!10,      
    colframe=gray!80,     
    title=Conclusion for RQ2
]
\method achieved an average number of repairs of \textbf{2.04} across both models, with a best average of \textbf{1.17} repairs observed on DeepSeek, evincing \method’s efficiency in repairing risk-prone configurations. 
\end{tcolorbox}

\section{Related Work}\label{Sec: related work}

Configurations play a crucial role in Cyber-Physical Systems (CPS). Prior studies in software engineering show that misconfigurations are a major cause of system failures \cite{lian2025large}, and numerous approaches have been proposed to detect or repair misconfigurations through static analysis \cite{rahman2023security, wang2023conftainter, timperley2022rosdiscover}, search-based methods \cite{valle2023automated, valle2025industrial}, or data-driven anomaly detection \cite{fellner2024pilot, fellner2024data}. 

In the context of unmanned aerial vehicles, existing work primarily aims at detecting misconfigurations rather than repairing them. For example, Ma et al. \cite{ma2023fuzz} proposed a fuzz testing approach based on evolutionary algorithms to generate configuration combinations that can lead to UAV unstable states. Chang et al. \cite{chang2023fuzzing} proposed APFuzzer, a state-guided fuzzing system that uses a Quality-Diversity Enhanced Genetic Algorithm (QDGA) to detect incorrect configuration parameters causing abnormal drone flight states. Later, Chang et al. \cite{chang2024low} proposed LDFuzzer, a low-cost, state-guided fuzzing system designed to search incorrect configuration parameter values in drone control systems. Purandare et al. \cite{purandare2024exploration} explored failures in small Uncrewed Aerial Systems (sUAS) controller software by leveraging software product line engineering to systematically analyze the configuration parameter space, uncovering failures caused by single parameter value changes and all pairs of changes. Cui et al. \cite{cui2024learning} proposed a machine learning-based approach to detect incorrect control parameter configurations in Robotic Aerial Vehicles (RAVs) that may cause operational issues. 

Besides detecting improper configurations, there is still some work that focuses on anomaly detection. For instance, AI Islam \cite{al2024real} proposed methods for detecting and diagnosing flight anomalies. Tan et al. \cite{tan2025runtime} proposed RADD (Runtime Anomaly Detection for Drones), an integrated approach combining rule mining and unsupervised learning to detect runtime anomalies in drones. Shar et al. \cite{shar2022dronlomaly} proposed DronLomaly, a deep learning-based log analysis approach for real-time detection of anomalous drone behaviors that may lead to physical instability or safety risks. Minn et al. \cite{minn2024dronlomaly} developed an automated tool (named DronLomaly) for runtime anomalous behavior detection of DJI drones’ flight missions based on flight logs.  
In addition, many other works use fuzzing methods \cite{ma2023fuzz, ma2024dynamic, malviya2025fuzzing} to detect bugs and vulnerabilities in Drones.

In summary, existing research predominantly focuses on identifying misconfigurations or anomalous behaviors. Current approaches either expose risk-prone configurations \cite{han2022control, han2024range, ma2023fuzz, chang2023fuzzing, chang2024low, purandare2024exploration, cui2024learning}, or detect anomalous behaviors \cite{al2024real, tan2025runtime, shar2022dronlomaly, minn2024dronlomaly, titouna2020online, li2024zero, kumar2024building}, but they lack providing automated mechanisms to restore system stability once a problematic configuration is detected. Moreover, prior studies have not explored the use of LLMs as a reasoning engine to analyze configuration–state relationships and further produce actionable misconfiguration fixes. To bridge this gap, we propose \method, enabling LLM-driven diagnosis and repair of configuration-induced drone anomalies.

\section{Experiences and Lessons Learned}\label{Sec: lessons learned}

\textbf{Model quality and token resources matter.} During our preliminary experiments, we observed a performance gap between LLM responses made using paid token resources and those relying on free token quotas. When using the free quota, the model occasionally produced less accurate configuration correction, resulting in more repair iterations before achieving a stable flight state. This suggests that the responsiveness and reasoning accuracy of our LLM-based repair mechanism are closely dependent on both model capacity and computational resources. Therefore, we recommend using advanced LLMs and sufficient token resources when applying \method to ensure timely responses and better repair effects.

\textbf{Necessity of embedding configuration constraints in prompt.} 
In the pilot experiments, we initially designed prompts that only included anomaly type and the corresponding configuration, without providing the official recommendations from the developers. Under such incomplete guidance, the LLM often generated less effective repairs. When we added constraints from the official documentation, the LLM produced more effective fixes. This suggests that configuration constraints are essential when applying LLMs to configuration repair, and providing explicit configuration boundaries makes LLMs more capable of producing effective repairs, thereby reducing repair iterations.

\textbf{Strengthening configuration management is essential for flight safety.}
Our experimental results indicate that the vast majority of misconfigurations we experimented with triggered anomalous flight states, and a substantial portion even resulted in crashes. This finding highlights that pre-flight configuration validation plays a critical role in preventing catastrophic failures. Effective configuration checks, which are the first line of safeguard for flight safety, should be strictly performed before flight missions.

\section{Conclusion and Future Work} \label{Sec: conclusion}

In response to the robustness degradation caused by risk-prone configurations in drone flight control systems, we proposed \method, an LLM-based method, which detects anomalies caused by risk-prone configurations and repairs them in real time through an iterative monitoring-repair mechanism. To evaluate \method, we conducted experiments on ArduPilot flight control software using a dataset comprising 1,421 misconfiguration cases and two representative large language models. The experimental results indicate that \method attained an average repair success rate of 90\% and an average number of repairs of 2.04 across two models, demonstrating both the effectiveness and efficiency of the proposed method. 

Future research will focus on the generalizability of \method. In particular, we plan to extend our method beyond drone flight control and explore automatic, real-time mitigation of improper configuration induced instabilities in other CPS, such as smart manufacturing systems and underwater robotic platforms. By adapting \method to different control architectures and operational environments, we aim to develop a unified and scalable solution for ensuring configuration reliability across diverse CPS domains.

\bibliographystyle{unsrt}  

\bibliography{references}

\end{document}